\documentclass{emulateapj}
\usepackage{natbib}
\begin{document}

 \title{A 500 Parsec Halo Surrounding the Galactic Globular NGC 1851}

\author{Edward W.~Olszewski, \altaffilmark{1,6} Abhijit Saha \altaffilmark{2,6}, Patricia Knezek \altaffilmark{2,6}, Annapurni Subramaniam \altaffilmark{3,6},\\Thomas de Boer \altaffilmark{4,6}, and Patrick Seitzer \altaffilmark{5}}
\email{eolszewski@as.arizona.edu}
\altaffiltext{1}{Steward Observatory, The University of Arizona, Tucson, AZ}
\altaffiltext{2}{National Optical Astronomy Observatory, Tucson, AZ}
\altaffiltext{3}{Indian Institute of Astrophysics, Bangalore, India}
\altaffiltext{4}{Kapteyn Astronomical Institute, University of Groningen, The Netherlands }
\altaffiltext{5}{Dept.~of Astronomy, University of Michigan, Ann Arbor, MI}
\altaffiltext{6}{Visiting astronomer, Cerro Tololo Inter-American Observatory, National Optical Astronomy Observatory, which are operated by the Association of Universities for Research in Astronomy, under contract with the National Science Foundation.}
\begin{abstract} 

Using imaging that shows four magnitudes of main sequence stars, 
we have discovered that the Galactic globular cluster NGC~1851 is
surrounded by a halo that is visible from the tidal radius
of 700 arcsec (41 pc) to more than 4500 arcsec ($>$250 pc).
This halo is symmetric and falls in density as a power law 
of $r^{-1.24}$. It contains approximately 0.1\% of the dynamical
mass of NGC~1851. There is no evidence for tidal tails.
Current models of globular cluster evolution do not explain
this feature, although simulations of tidal influences on
dwarf spheroidal galaxies qualititively mimic these results.
Given the state of published models it is not possible to decide
between creation of this halo from isolated cluster evaporation,
or from tidal or disk shocking, or from destruction of a dwarf
galaxy in which this object may have once been embedded.

\end{abstract}

\section{Introduction}

While it is clear that globular clusters must evaporate, and that
globulars in a tidal field can also lose mass through their Lagrangian
points, the exact state of disruption of many Galactic globulars is
less certain.  
Much of the search for tidal structure (and for the fundamental
parameters of globulars, for that matter) has been done with star
counts from photographic plates (Grillmair et al.~1995, Leon et
al.~2000, Illingworth \& Illingworth 1976, Peterson and King 1975, and
references therein).  We give two examples of problems with the
photographic datasets: NGC 288, for instance, has been claimed to have
``extratidal structure'' (Leon et al.~2000), while recent kinematic
work finds no evidence for a coherent velocity gradient in the
direction of the tidal tails (Kiss et al.~2007).  Similarly, Da Costa
\& Coleman (2008) use kinematics to put a limit of 0.7\% of the mass
of $\omega$ Cen between 1 and 2 tidal radii.

SDSS, HST, and mosaic cameras on large telescopes have allowed
new insights into the structure of globular clusters.
Recently several clusters have been shown
to have spectacular tidal tails outside the tidal radius: Palomar 5
(Odenkirchen et al.~2001, Rockosi et al.~2002, 
Odenkirchen et al.~2003) and NGC 5466 (Belokurov et al.~2006, Grillmair \& Johnson
2006) are
prototypes of this class of globulars.  In addition, Lauchner et al.~(2006)
use SDSS to claim tidal tails in NGC 5053, though most of the stars
are confined within the tidal radius.
 Sohn et al.~(2003) use CFHT Mosaic imaging to discover ``tentative
detections of tidal halos'' around Pal 3 and 4.
The typical such cluster is
``loose'', with a relatively small ratio of tidal to core radius.
Even those clusters that are convincingly being destroyed show the
onset of two tidal tails at a relatively small distance from the
tabulated cluster tidal radius.

It has been known for more than 40 years that the main
bodies of globular clusters can be described by King (1966) models.
Limiting (tidal) radii from these King-model fits range from 1.6 pc
(NGC 6544, galactocentric distance 5.3 kpc) to 214 pc (NGC2419,
galactocentric distance 91.5 kpc), with a median of 36 pc (derived
from Harris 1996).  90\% of all Galactic globulars have radii of less
than 100 pc (Harris 1996). 

McLaughlin \& van der Marel (2005) have shown that Wilson (1975)
models of the stellar distributions of globulars in the Milky Way, the
Magellanic Clouds and the Fornax dwarf better describe the outer
structures of thse clusters than do King models. McLaughlin and van
der Marel (2005) homogeneously re-fit King, Wilson, and power-law
models to 153 of these clusters. In general, the new King tidal radii
are similar to those in the online Harris catalog, while the Wilson
tidal radii are larger.  Typically, the Wilson tidal radii are a
factor of 2.5 larger than the King tidal radii (McLaughlin and van der
Marel 2005), while for NGC 1851 the Wilson radius is a factor of 6.8
larger (McLaughlin and van der Marel's King tidal radius for
NGC 1851 is 6.7 arcmin as opposed to 
11.7 in the Harris compilation, with a Wilson tidal radius of
45 armin).  We will use the King tidal radii in this paper simply because
they are better known, and since in the case of NGC 1851 we believe
that the fundamental data need improvement (star counts only extend to
10 arcmin from the center) to determine a realistic limiting radius.
For the rest of this paper, we will use the phrase ``tidal radius'' to
mean the King tidal radius tabulated in the online Harris catalog
unless stated otherwise. Both the King tidal radius and the Wilson
tidal radius are typically beyond where the star counts end,
and can be poorly constrained in specific globulars.

What is the outer structure of a typical globular? Until recently, NGC
1851, the subject of this paper, would have been such a cluster: its
absolute magnitude of $-8.3$ is 1 mag brighter than the median
Galactic globular, and the tidal radius of 41 pc is very close to the
the median.  NGC 1851, however, has recently been shown (Milone et
al.~2008), using HST imaging, to be a most remarkable object. Above
$\sim$m$_{F606W}$$=$19, there are two distinct main-sequences (MS)
and lower subgiant branches (SGB) visible, making it one of a small
number of clusters with dual stellar populations.  Milone et
al.~(2008) see no relative changes in the 2 subgiant populations out
to a distance of 1.75 arcmin. Zoccali et al.~(2009) study the radial
extent of the two subgiant branches from ground-based data, and claim
that the fainter sequence dies away 2.4 arcmin from the center (tidal
radius is 11.7 arcmin), though Milone et al.~(2009) refute the Zoccali
work.  Dinescu et al.~(1997) derive the absolute proper motion for NGC
1851, which when coupled with the large (320.5 km/s) heliocentric
radial velocity and current distances (R$_\odot$$=$12.1 kpc,
R$_{gc}$$=$16.7 kpc, X$=$ $-$4.3 kpc, Y$=$ $-$8.9 kpc, Z$=$ $-$6.9
kpc), gives an orbit with period $\sim$0.4 Gyr, apogalacticon distance
$\sim$30 kpc, and perigalacticon distance $\sim$ 5 kpc. This orbit
passes through the disk of the Milky Way five times per Gyr at
distances ranging from 5 to 30 kpc (orbital calculations graciously
performed by Piatek (2009)).  According to the theoretical papers
cited in the discussion, disk shocking and tidal shocking should be
important for NGC~1851, and a tidal tail should be seen. Leon et al
(2000, their figure 10) claim to see symmetric structure to about 20
arcmin radius (about 1.7$\times$ the tidal radius) and then
``extensions which are likely tracing the orbital path...''  Leon et
al (2000) azimuthally average background-count subtracted data and fit
a power law to radially distant points, finding a power law slope for
NGC 1851 of $-$0.98, similar to what we find in Section 2. Our new
observations contradict these claims: we see no tidal tails and do not
see discrete structures in common with the photographic counts, but
show smooth extended structure.

\section {Observations and Data Reduction}

\begin{figure*}
\begin{center}

\includegraphics[width=16cm,angle=90]{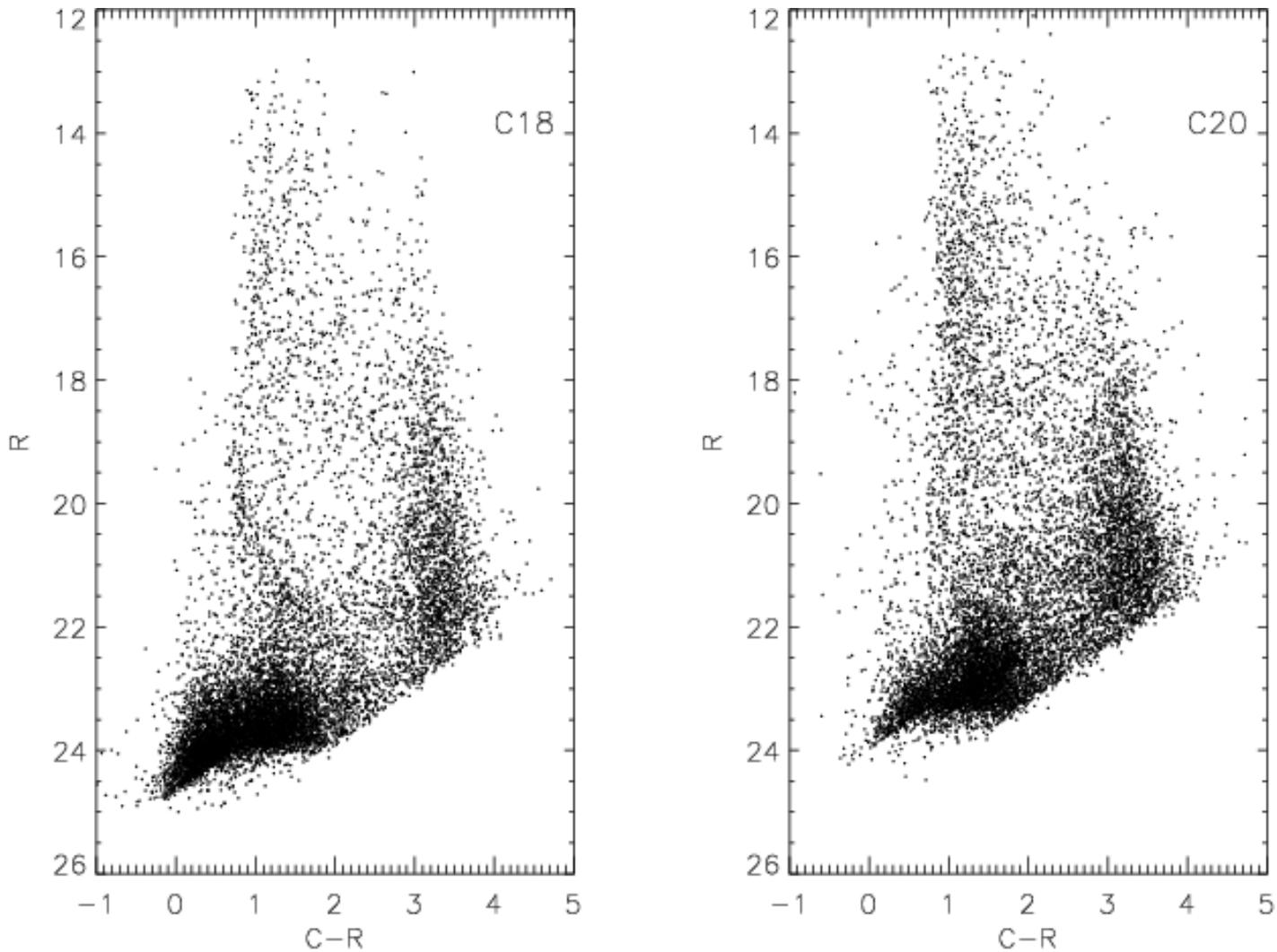}
\caption{R versus C-R cmds for control fields C20 and C18 (near NGC 1851)
from another project. Note the wall of F stars and the ``extra'' main
sequence in C18.}
\label{fig:Washington}
\end{center}
\end{figure*}

We discovered that NGC 1851 has extended structure while observing a
control field, centered 3.5 tidal radii from NGC 1851, for a project
to probe the outer structure of the LMC.  The R versus C$-$R (C is
Washington C (Canterna 1976), centered at $\sim$4000 \AA)
color-magnitude diagram (CMD) of this field C18 (l=245, b$=-$35) is
compared to C20 (l=245, b$=-$25), 10 degrees away, in
Figure~\ref{fig:Washington}. A main sequence is obvious in C18 from
R$\sim$18.5 to R$\sim$22.  C20, at lower latitude, has more stars
overall, but does not show this additional feature. This feature is
most likely a main sequence of an object at a single distance, and
corresponds exactly to the main-sequence locus in published NGC 1851
main body photometry (e.g., Figure 10 in Walker 1992), after a
reasonable color transformation.

We therefore began to map out this main sequence by acquiring nine
fields outlining the cluster from $\sim$1.0--$\sim$6.5 tidal radii.
Exposures were 300 seconds in V and 600 seconds in B, using the CTIO
4m Blanco telescope with the Mosaic II CCD array. Each field is
36$\times$36 arcmin on a side. Table~\ref{table:obsinfo} gives a diary
of observations of the BV fields.

\begin{deluxetable*}{lllr}

\tablecolumns{4}

\tablecaption{Log of BV observations of NGC~1851 \label{table:obsinfo}}

\tablehead{
\colhead{Field}& \colhead{RA J2000.0}& \colhead{Dec J2000.0}& \colhead{M.J.D.~of V data}}\\

\startdata

NGC~1851 inner&       $05:15:06.8$& $-$$40:26:47$&    54825.20172858  \\ 
NGC~1851 E&           $05:16:57.1$& $-$$40:02:51$&    54715.40055975 \\
NGC~1851 W&           $05:11:16.0$& $-$$40:03:31$&    54823.28117295 \\
NGC~1851 N&           $05:14:07.1$& $-$$39:29:57$&    54822.33423601  \\
NGC~1851 S&           $05:14:07.2$& $-$$40:35:50$&    54715.38609743 \\
NGC~1851 NE&          $05:16:59.9$& $-$$39:29:52$&    54823.32000866   \\
NGC~1851 NW&          $05:11:15.0$& $-$$39:29:54$&    54823.33312843  \\
NGC~1851 SE&          $05:16:59.3$& $-$$40:35:47$&    54823.29429358  \\
NGC~1851 SW&          $05:11:15.0$& $-$$40:35:51$&    54823.30709821  \\

\enddata

\end{deluxetable*}

Data were reduced using IRAF's{\footnote{ IRAF is distributed by the
 National Optical Astronomy Observatory, which is operated by the
 Association of Universities for Research in Astronomy (AURA) under
 cooperative agreement with the National Science Foundation.}  MSCRED
 package and a set of homegrown IDL routines (Saha et al.~(2010).
 Photometry is then done with a privately modified (by Saha) version
 of DoPHOT (Schechter et al.~1993) and with a set of post-DoPHOT
 programs written by one of us (Saha).  Two aspects of the reduction
 are relatively uncommon: first, we create single resampled images
 from the 8-CCD mosaics as input to DoPHOT. Our extensive testing
 shows that the photometry is unaffected by our resampling of the
 images (Saha et al.~2010). Second, DoPHOT was run with a
 non-variable PSF.  However, the real PSF does vary with position on
 the field-of-view, so the PSF fitted magnitudes have consequent
 position dependent trends of several percent. These were corrected by
 obtaining aperture magnitudes for the brighter high S/N stars (which
 do not vary with position), and comparing the PSF fitted magnitudes
 to the aperture measured values, and thus deducing the position
 dependence of the fitted magnitudes. The derived corrections as a
 function of position on the field-of-view were then applied for all
 stars.  This procedure is more robust for these data than using a
 variable PSF.  These extra steps also allow for corrections due to
 tilt and piston and defocus as a function of position on each
 original CCD. There is demonstrably little chip-to-chip photometry
 shift remaining after using this technique. The small chip-to-chip
 and field-to-field photometry errors from these reduction procedures
 keep the photometry from wandering in zeropoint and scale, and permit
 the large-scale mapping described here.

The nine fields are combined by using stars in common between frames
to adjust the instrumental magnitudes to a common scale, the scale of
our innermost field. The photometric zeropoint of the innermost field
comes from comparison of published NGC 1851 photometry (Walker 1992).
In other words, photometry within $\sim$1 tidal radius was forced
to that of Walker (1992), but the main sequence outside that radius 
is bootstrapped to the magnitudes so obtained in the innermost field
(field 2 shifted by using overlap region of fields 1 and 2, field 3
shifted by using overlap of fields 1 and 2 against it, etc.)
Thus any excursions of main sequence structure over the extended
field are preserved, if they exist.
The upper plot of Figure~\ref{fig:Full} shows the CMD of the entire area imaged, including the small region inside 1 tidal radius. We see a
clear main sequence and lower subgiant branch, the wall of thick disk
and halo turnoff stars, the wall of K and M stars, and the blob of
faint galaxies.  The lower plot shows only stars more than 2000 arcsec
from the cluster center (more than 2.85 tidal radii). In both CMDs
there are two turnoffs and lower subgiant branches (Figure~\ref{fig:histo}), extending the Milone et 
al.~(2008) HST result to extraordinarily large radii. This result
contradicts the Zoccali et al.~(2009) claim that the fainter SBG
disappears 2.4 arcmin from the cluster center, and is consistent
with the recent Milone et al.~(2009) result which follows the two
subgiant branches out to at least 8 arcmin.

\begin{figure*}
\begin{center}

\includegraphics[width=16cm]{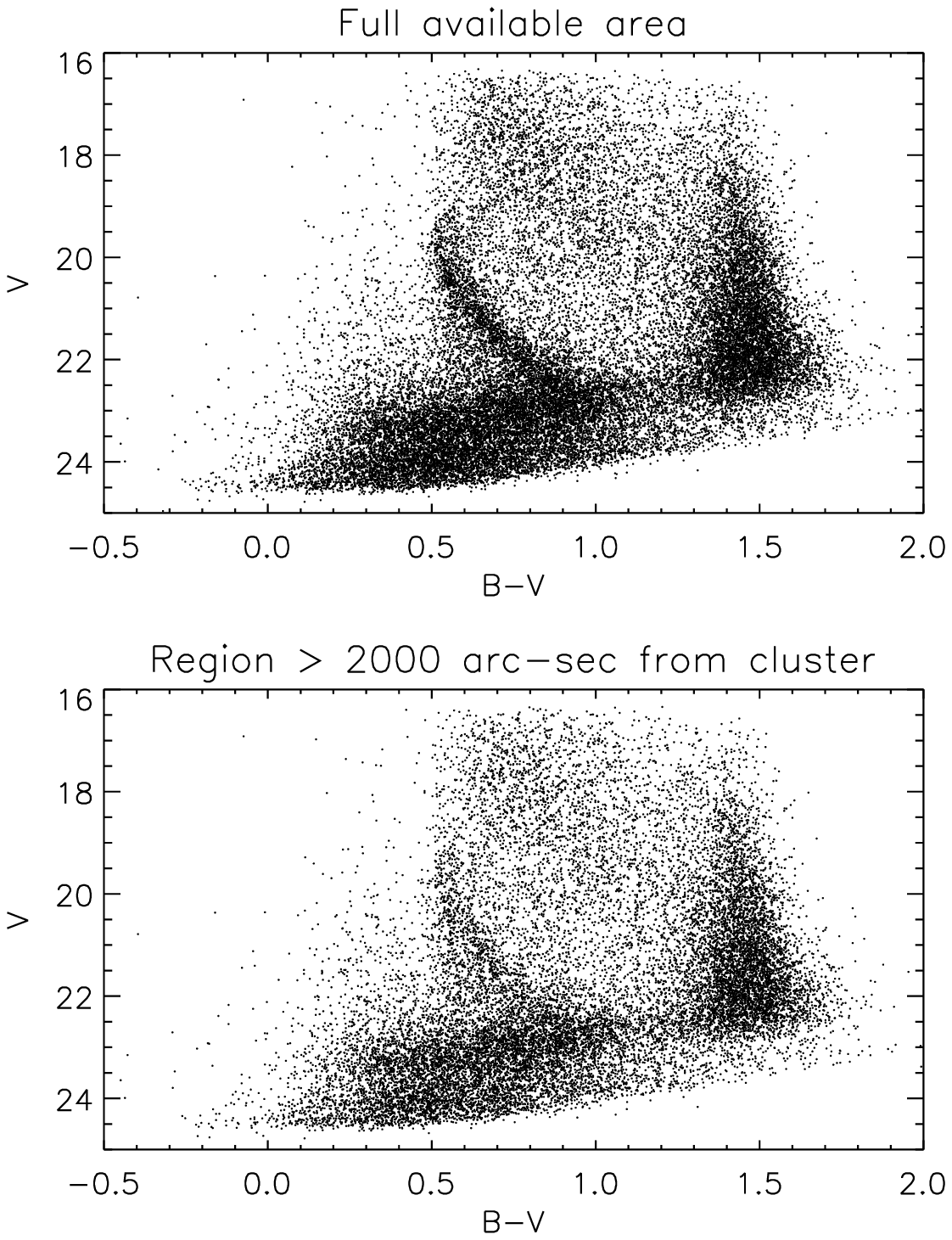}
\caption{V versus B-V cmds for two regions surrounding NGC 1851.
Top diagram is of all acquired data. Bottom figure is of
a region more than 2.85 tidal radii from NGC 1851. In both cases
the main sequence is identical to that of the globular, and the double
subgiant branches are visible.}
\label{fig:Full}
\end{center}
\end{figure*}

\begin{figure*}
\begin{center}

\includegraphics[width=16cm]{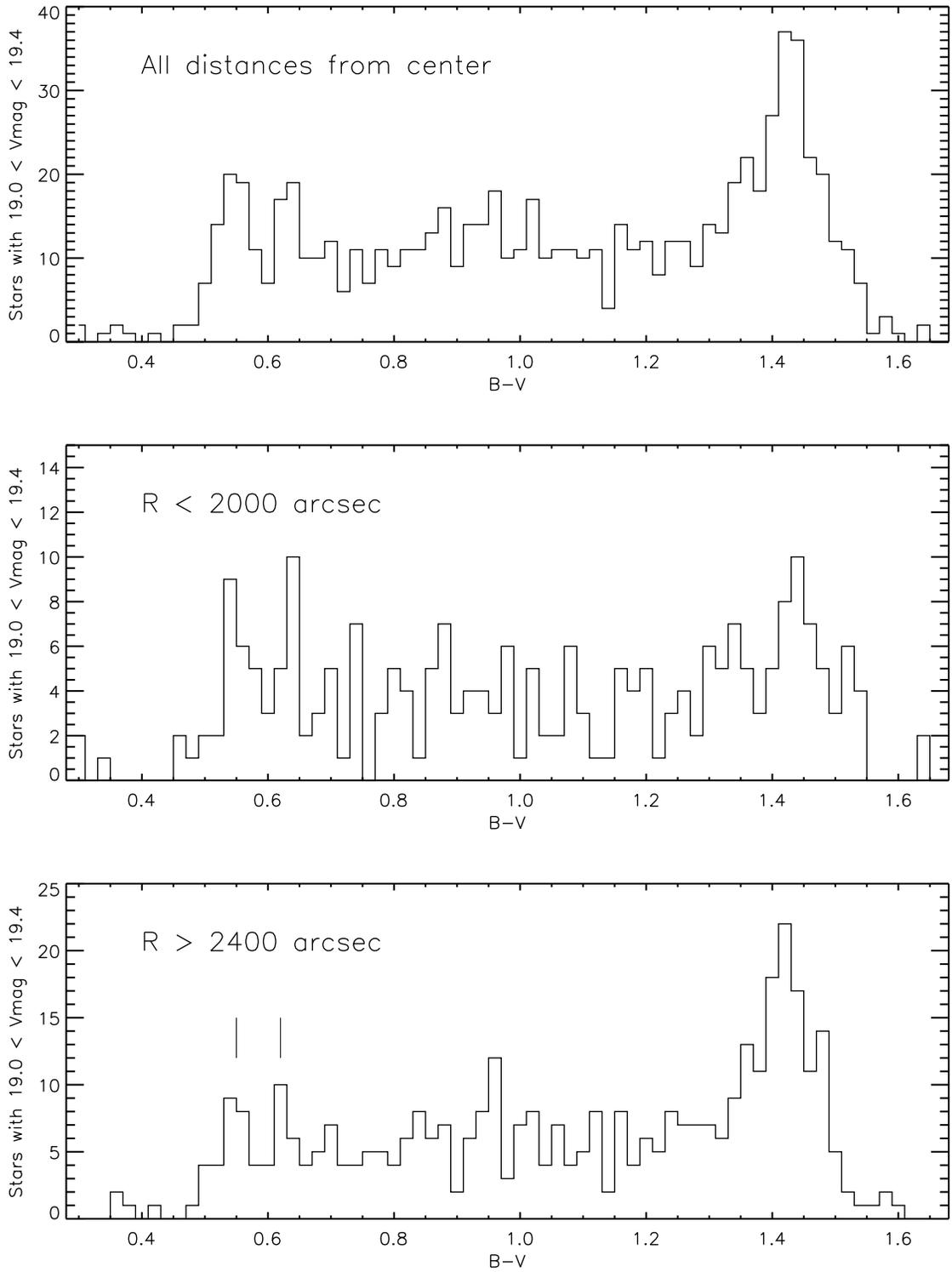}
\caption{
Histograms of all stars with 19.0 $<$ V $<$ 19.4, in B$-$V bins of 0.02,
to show peaks in the CMD. The features at B$-$V$=$0.55 and 0.63 are
the two stellar populations in NGC 1851. The peak at about B$-$V$=$1.4 
is made up of the foreground K and M stars. Top: Histogram of all
observed stars; Middle: Histogram of all stars closer than 2000 arcsec (117 pc); 
Bottom: Histogram of all stars more than 2400 arcsec (140 pc) from
the center to about 4000 arcsec (230 pc). The two peaks of interest
are marked with tickmarks in the bottom plot.
}
\label{fig:histo}
\end{center}
\end{figure*}

We define a region on the CMD that contains the main sequence and
lower subgiant branch.  This region is a band that follows the ridge
line of the MS and SGB and is 0.15 mag wide in (B-V) (see
Figure~\ref{fig:ridge}).  Background estimates come from regions
centered 0.15 mag redward and blueward of this box, and by scaling the
C20 control field in Figure~\ref{fig:Washington}.  In
Figure~\ref{fig:position} we plot the positions of stars in our MS and
SGB regions readily distinguished in Figure~\ref{fig:Full}. We also
plot the distribution of stars in the two background regions of the
CMD.  Aside from a density peak of non-cluster stars at large
distances to the East, there are no obvious
trends. Figure~\ref{fig:angular} shows the angular distribution of
stars outside the tidal radius, in 15$^\circ$ bins, of stars in the
main-sequence region of the CMD (dashed), in the two background
regions of the CMD (dotted), and in the total number of NGC stars
(solid), which is the counts in the dashed histogram minus half of the
dotted. Counting errors for the solid curve are $\pm$7 on average,
which encompasses most of the bins.  There is no evidence, therefore,
for tidal tails or angular structure on the distribution of NGC 1851
stars.
 
Counting stars on the main
sequence, and counting stars to the red and blue of the main sequence
to use as background estimations, gives the density plot of
Figure~\ref{fig:slope}.  We see steeply declining counts from about
500 arcsec to about 1000 arcsec (1.3$\times$ the cataloged tidal radius),
with a very shallow decline out to the limits of our imaging, more
than 4000 arcsec. Fitting a straight line to this halo, beyond 1200
arcsec, gives $\log(counts)\propto \log(r)^{-1.24}$, with 1-$\sigma$
errors on the power law of $-0.58$ and $-1.9$. Figure~\ref{fig:linear}
shows the density plot in linear units to underscore the large extent
of the NGC~185 halo. While our discovery of stars out to 67 arcmin
may not seem extreme if one believes the Wilson tidal radius of 45 arcmin
for NGC 1851, we note that the King model fit using the code of
McLaughlin and van der Marel (2005) gives no stars beyond 5.8 arcmin,
and the Wilson model gives no stars beyond 43 arcmin. We can still
see the main sequence by eye (Figure~\ref{fig:Full}) at radii
extending beyond the (very uncertain)Wilson tidal radius. 

\begin{figure*}
\begin{center}
\includegraphics[width=16cm]{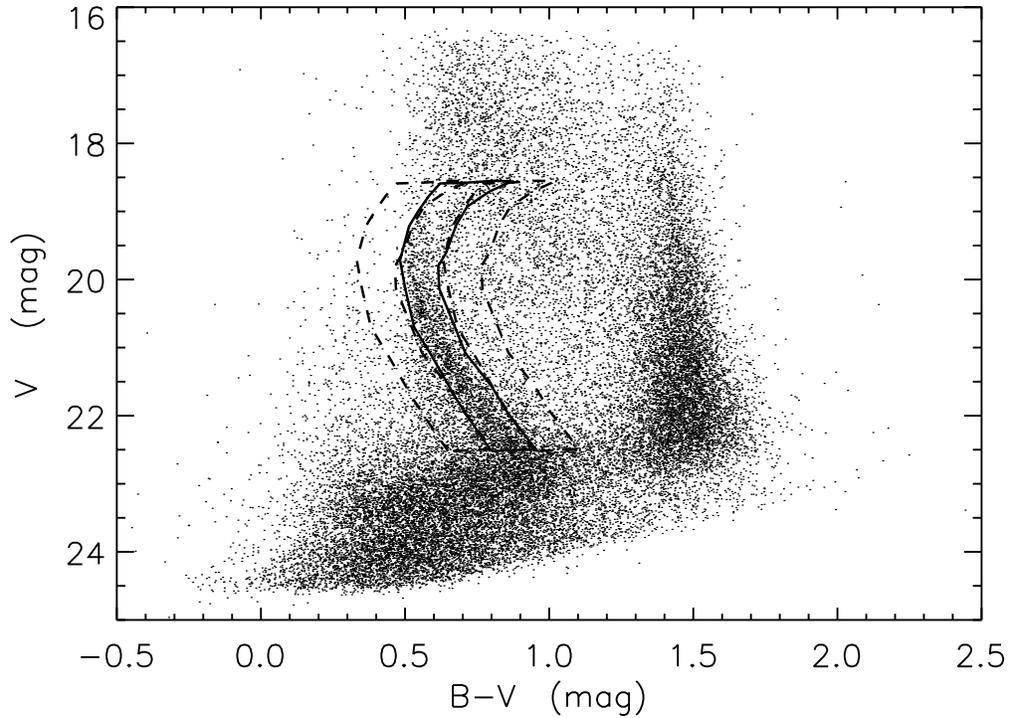}
\vspace{-4in}
\caption{CMD of NGC 1851 halo with our main-sequence and subgiant regions,
and the two off-sequence background regions overlaid. Each region
is 0.15 mag wide in B-V. There is a very small overlap at
V=18.5.}
\label{fig:ridge}
\end{center}
\end{figure*}

\begin{figure*}
\begin{center}
\includegraphics[width=9.5cm,angle=90]{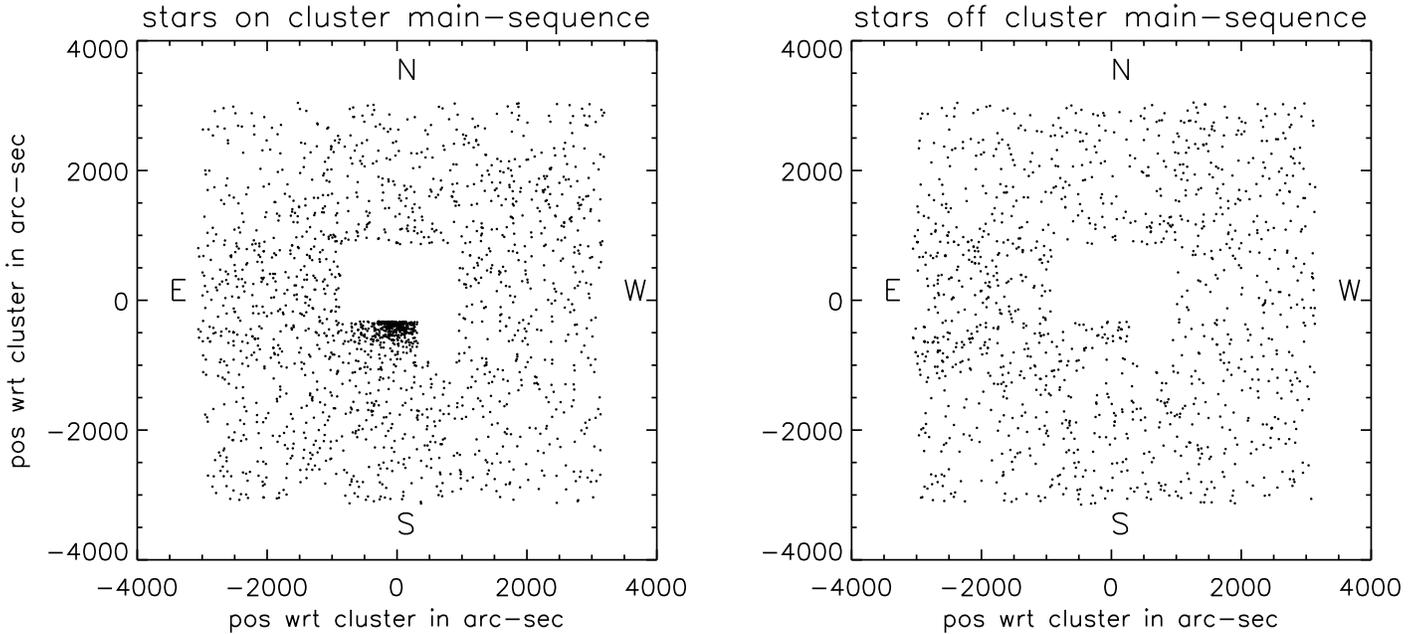}
\caption{ Left: positions of all stars that meet our criteria
for being on the observed main sequence and subgiant branches.
Regions interior to 1 tidal radius  (700 arcsec) 
were not observed except for one small
region to the south.
The extreme plotted points are
more than 6 tidal radii from the center. Right: the same plot for the 
distribution of stars in the two background regions. An excess of field
stars can be seen to the East.}
\label{fig:position}
\end{center}
\end{figure*} \begin{figure*}

\begin{center}
\includegraphics{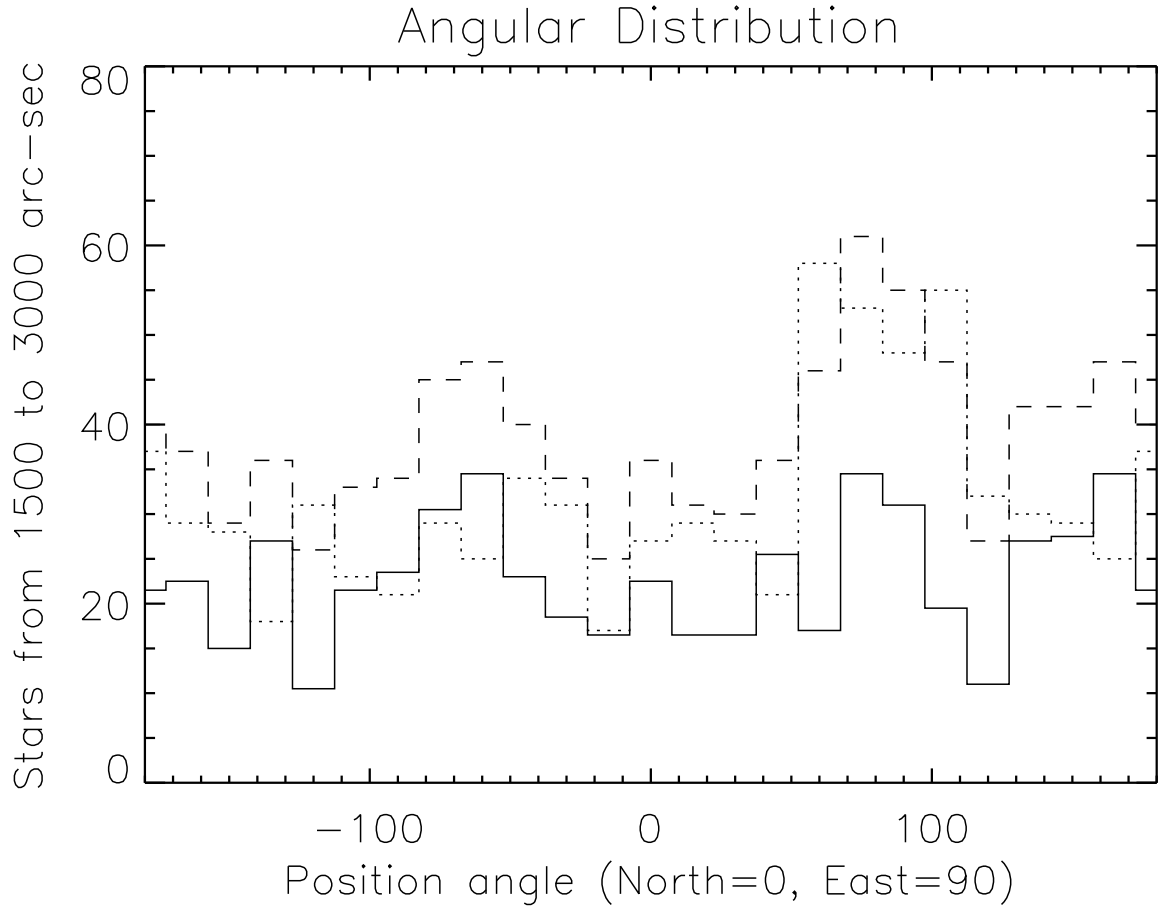}
\vspace {10pt}
\caption{ The angular distribution of stars as a proof of the lack
of tidal tails. Dashed histogram: all stars outside of $\sim$1 tidal radius,
in the main sequence region of Figure~\ref{fig:ridge}. Dotted histogram:
all stars in the two off-cluster positions (twice the area in
the CMD) shown in Figure~\ref{fig:ridge}.
Solid histogram: the difference ``dashed'' minus 0.5$\times$''dotted'' (there
are two off-main-sequence regions). This is the
total number of NGC 1851 stars in the main sequence region.
The average is about 25$\pm$7 for the solid histogram, showing
the lack of statistically significant features in the azimuthal distribution.
}
\label{fig:angular}
\end{center}
\end{figure*}

\begin{figure*}
\begin{center}
\includegraphics[width=16cm]{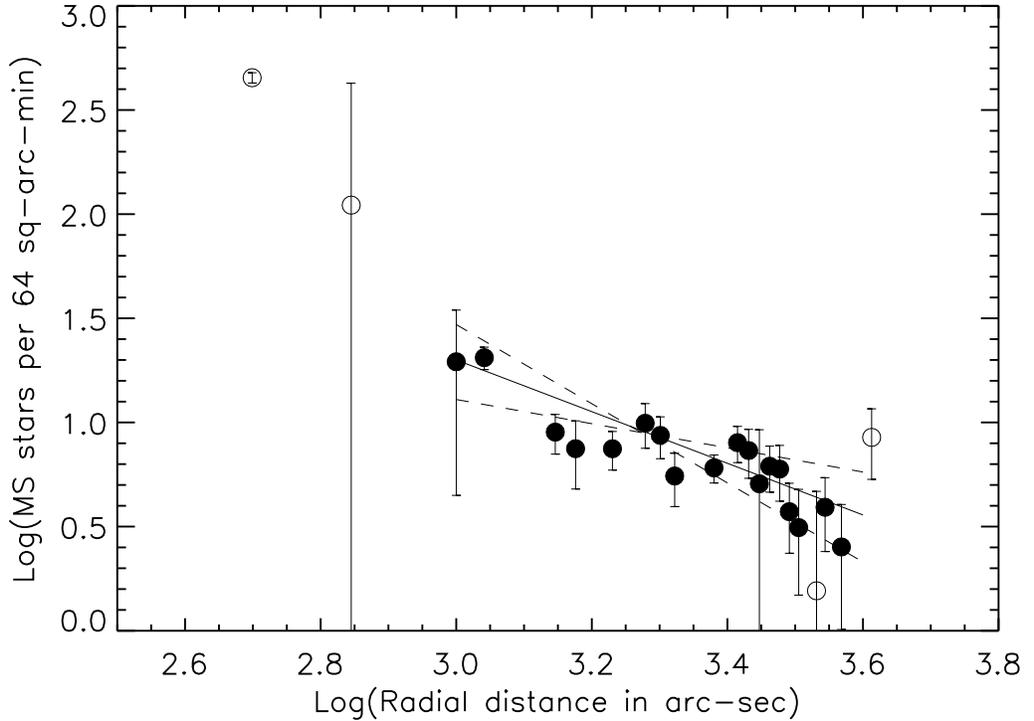}
\vspace{-4in}
\caption{Density plot of stars surrounding NGC 1851. Note the Y scale
of log of stars per 64 arcmin$^2$.Open circles are regions in radius with
very incomplete coverage. Filled circles 
between 1200 and 4000 arcsec were used for power-law fits.
Solid line is the best fit ($\chi^2=0.9$) of r$^{-1.24}$; dashed  lines
are the 1-$\sigma$ limits of r$^{-1.9}$ and r$^{-0.58}$.}

\label{fig:slope}
\end{center}
\end{figure*}

\begin{figure*}
\begin{center}
\includegraphics[width=16cm]{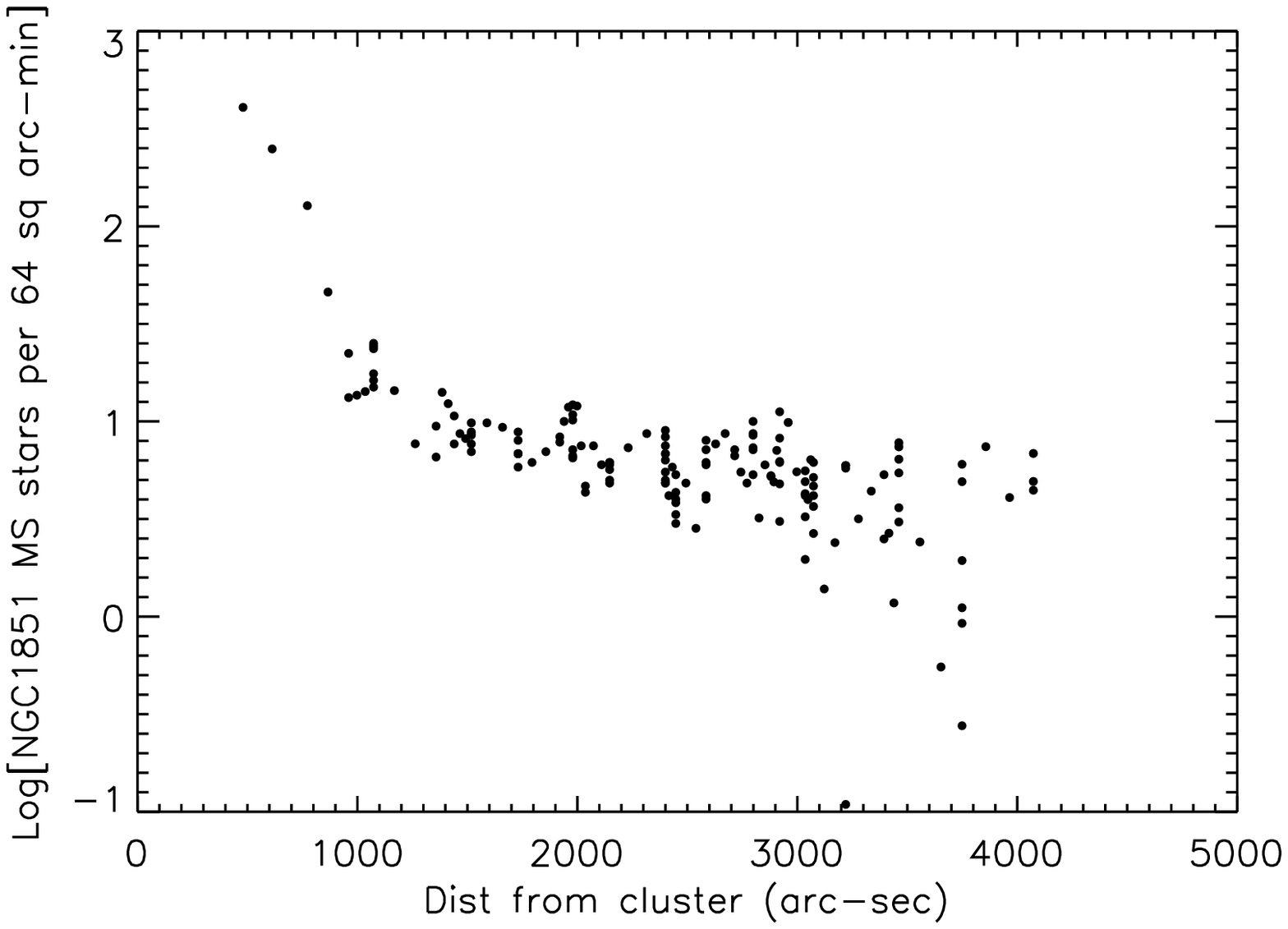}
\caption{Density plot of stars surrounding NGC 1851. The x axis is now
linear in radius. }
\label{fig:linear}
\end{center}
\end{figure*}

\section {Discussion}

Unlike the case of the Pal 5 tidal tails, this set of extratidal stars
does not make up a large fraction of the current mass of the cluster.
For a region exterior to 13 arcmin from the center, we count 1647
total stars in the main sequence region of Figure~\ref{fig:ridge}.
Using off-main-sequence regions defined in Figure~\ref{fig:ridge} and
in a scaled version of control field C20, we measure that 699 of those
stars are field stars.  The net number of stars outside 13 arcmin is
therefore 948$\pm$40.5, or 58\% of the total. Despite the extremely
low surface brightness of ~$\sim$32 mag arcsec$^{-2}$, the main
sequence and the SGB remain vivid.

If the typical mass on this main sequence is 0.5 M$_{\odot}$, this
halo of 500 M$_{\odot}$ measures 0.1\% of the dynamical mass of NGC
1851. NGC 1851 is therefore embedded in a very low-mass halo 500pc
across, or ``dwarf spheroidal size''.  This halo also contains two
populations, as seen in the two main sequences, of unknown origin.

The state of modelling of stars outside the King tidal radius and {\it not} in
tidal tails, is unsatisfying.  Observationally, as pointed out in
Odenkirchen et al.~(2009), the expected signatures of tidal mass loss
are not as clear as hoped.  Gnedin and Ostriker (1997) model the
destruction rates of globulars, but do not provide detailed spatial
stellar distributions.  A number of papers (e.g., Spitzer 1940; Gnedin
\& Ostriker 1997) show that mass loss, because of internal 2-body
interactions that force lower-mass stars slowly to the periphery of
the cluster, changes the mass-function from center to edge of the
cluster.  Combes et al.~(1999) find that unbound stars (not in the
tidal tails) change slope at the tidal radius, to a surface density
slope of $-$3. Combes et al.~show a time sequence of their model m2
(their table 1 and figure 14). m2 has a tidal radius of 60 pc. Even
after significant mass loss the inner symmetrical part stays at about
60 pc radius, with the tidal tails emanating from the tidal radius
area.  Fukushige \& Heggie (2000)
use simulations to show that stars inside the tidal radius with energies above the
escape energy can leave the cluster on a similar timescale
to stars leaving due to 2-body interactions, approximately a Hubble time.
This simulation might come closest to mimicking a cluster like NGC 1851,
except for the complications due to the orbital eccentricity
of NGC 1851.
Mass loss in a slightly different context, that of dwarf
spheroidal galaxies, has been studied by Johnston et al.~(1999) and by
Pe{\~n}arrubia et al.~(2008).  Johnston et al (1999), in a study of the
mass loss rates from dwarf spheroidal galaxies, point out the
existance of a change of slope of the stellar density (star counts) at
the radius where now-unbound stars become important. They derive a
much shallower slope for the region outside the tidal radius, $-1$ or
so, in contrast to that of Combes et al. 

Typical observational cases might be those of NGC 2419 and M92. Ripepi
et al.~(2007), see main sequence stars in NGC 2419 out to more than 1.2
tidal radii.  Lee et al.~(2003) show the radial profile for M92, with
power law fits of slope $\sim -1$ beyond the tidal radius, but with
symmetric structure out to perhaps 1.5$\times$ the tidal radius,
little evidence for tidal tails, and possible clumpy structure outside
the symmetric structure. This is qualitatively consistent with Testa
et al.~(2000) whose analysis of POSS data shows symmetric structure out
to perhaps 1.5-2$\times$ the tidal radius, plus more distant irregular
structure. Testa et al.~counted stars to 16th magnitude in photographic
F, while Lee et al.~counted stars to V=20.0 and to 23.5. Some of the
radially distant structures are in common to both studies, at least
qualitatively.  However, Drukier et al.~(2007) show that beyond the M92
tidal radius, almost no stars have the M92 velocity (their Figure 5).

In addition, good evidence for extratidal material in some globulars
has been overlooked.  Lee et al.~(2004, their Figure 5) show the CMD of
a control field at more than 2.5 tidal radii from NGC 7492 that
clearly shows the cluster main sequence.

Given the state of the modelling described above, it is impossible to
distinguish between creation of such a halo from 2-body internal
interactions, disk shock heating, tidal heating, destruction of a
primordial dwarf galaxy in which NGC 1851 was embedded, and initial
conditions. Nonetheless,
we briefly examine some of these.

Mass loss caused by Galactic tides is the obvious explanation for the
long tidal tails of objects such as Pal 5 (Odenkirchen et al.~2001;
Rockosi et al.~2002; Odenkirchen et al.~2003; Koch et al.~2004).  The Pal
5 tidal tails emerge from the cluster at about the tidal radius of 16
arcmin, or $\sim$100 pc (Odenkirchen et al.~2001, their Figure 2).  We
see no sign of tidal tails in the NGC 1851 halo out to the limit of
our survey, a radius of 250 pc. The orbit of Pal 5 (Scholz et al.
1998), with $R_{min}$ between 5 and 9 kpc, $R_{max}$ between 17 and 18 kpc,
and 
orbital period 0.4 Gyr, is qualitatively similar to that derived for
NGC 1851 above. Since Pal 5 is much fluffier than NGC 1851, we would
indeed expect that Pal 5 would be closer to dissolution than NGC 1851,
as it is.  Leon et al.~(2000) claim evidence for tidal tails tracing
NGC 1851's orbit, evident at distances of 15-40 arcmin (50-140
pc). Our CMDs contradict these claims.

Gnedin and Ostriker (1997) show that the evaporation time for NGC 1851
is very long, about 10 Hubble times. Gravitational shocking is also of
order the same value. If the halo of NGC 1851 is simply an evaporated
or shocked halo, there is qualitative consistency with Pal 5: Pal 5
(from their table 3) is quickly destroyed, while NGC 1851 is very
slowly destroyed. The models do not provide the information
necessary to compare our empirical halo with a model halo.

The studies of destruction of dwarf spheroidals (Johnston et al.~1999;
Pe{\~n}arrubia et al.~2008, their Figure 2) offer a possible explanation for
the structure of NGC
1851. The Pe{\~n}arrubia et al.~paper models a dwarf spheroidal as an NFW
dark-matter profile with a much more concentrated baryonic
component. NGC 1851 would be an even-more
concentrated nucleus in such a model. Pe{\~n}arrubia et al.~(2008, their
figures 2 and 3) show the effect of stripping away most of the mass of
the dSph. What remains is an object that can be fit to a King model,
but with a shallow halo outside the place where the crossing time
equals half the orbital time. One probably cannot extend these models
to the remnants of a nucleated dwarf, so we must appeal to n-body
experts to make the proper models for NGC 1851. The most striking
feature of NGC 1851 when compared to $\omega$ Cen, M54, and NGC 2419,
clusters that have at one time been hoped to be remnant nuclei of
nucleated dwarfs, is the small core radius, 0.2 pc, of NGC 1851.  NGC
1851 (Yong et al.~2009, Ventura et al.~2009 ), similarly to $\omega$ Cen, does have a large
spread in the elemental abundances C+N+O, making it different from
most globulars.

Of the 24 Galactic globulars more luminous than NGC 1851, the four
clusters with the largest current galactocentric distances all have
larger Wilson tidal radii than NGC 1851, and three have interesting
associations with stellar streams: NGC 1851 is on an elliptical orbit
reaching in to a galactocentric distance of 5 kpc and out to 30 kpc
(Section 1). NGC 2419 might be associated with the Virgo Stellar
Stream, which is on an extremely elliptical orbit (Casetti-Dinescu et
al.~2009).  NGC 5824 may be associated with a stellar stream (Newberg
et al.~2009).  NGC 6715 is M54, the core of the Sgr dSph.  The other
three luminous clusters with larger Wilson tidal radii than NGC 1851
are interior to the current position of NGC 1851, with NGC 6139 the
most extreme (R$_{GC}=1.7$ kpc and R$_{tidal(Wilson)}=676$ pc.) The
orbit of 47 Tuc is rather circular in the XY plane (R$_{apo}=7.3$ kpc,
R$_{peri}=5.2$kpc), while M3's orbit has an ellipticity of 0.42 
(R$_{apo}=13.4$ kpc, R$_{peri}=5.5$kpc)(Dinescu et al.~1999). A parameter
that relates
luminous globulars to large physical size might
therefore be some combination of short-period orbit, eccentric orbit, or
association with a stellar stream.

While the correct explanation for the halo of NGC 1851 is presently elusive,
the observational imaging result is clear and strong. Modern imaging
and reduction techniques can therefore tell if NGC 1851 is an
exceptional cluster and if the fundamental data for the
150 Milky Way globular clusters
need to be much improved and extended with wide-field CCD photometry.

We thank the CTIO mountain staff for their help in observing, both on
the mountain and from La Serena. We also thank Chris Smith for
computer, archive, wiki, and ``MOSOCS'' help.  Slawomir Piatek
performed the new orbital calculations for NGC 1851. Bill Harris introduced us
to the McLaughlin and van der Marel paper, and Dean McLaughlin
provided us with his fitting code and thoughtful comments
on the quality of tidal radius determinations. We thank the referee for
helpful comments. EO was partially
supported by NSF grants AST-0505711 and 0807498.

\end{document}